\documentclass[twocolumn,showpacs,preprintnumbers,amsmath,amssymb,superscriptaddress]{revtex4}

\usepackage{graphicx}
\usepackage{dcolumn}
\usepackage{bm}
\usepackage{braket}

\begin{document}


\title{Cu substitution effects on the local magnetic properties of Ba(Fe$_{1-x}$Cu$_{x}$)$_{2}$As$_2$:
a site-selective $^{75}$As and $^{63}$Cu NMR study}

\author{Hikaru Takeda}
\affiliation{Department of Physics and Astronomy, McMaster University, Hamilton, Ontario L8S4M1, Canada} 
\affiliation{Department of Physics, Graduate School of Science, Nagoya University, Furo-cho, Chikusa-ku, Nagoya 464-8602, Japan}
\author{Takashi Imai}
\affiliation{Department of Physics and Astronomy, McMaster University, Hamilton, Ontario L8S4M1, Canada} 
\affiliation{Canadian Institute for Advanced Research, Toronto, Ontario M5G1Z8, Canada} 
\author{Makoto Tachibana}
\affiliation{Department of Physics and Astronomy, McMaster University, Hamilton, Ontario L8S4M1, Canada} 
\affiliation{National Institute for Materials Science, Tsukuba, Ibaraki 305-0044, Japan}
\author{Jonathan Gaudet}%
\affiliation{Department of Physics and Astronomy, McMaster University, Hamilton, Ontario L8S4M1, Canada} 
\author{Bruce D. Gaulin}%
\affiliation{Department of Physics and Astronomy, McMaster University, Hamilton, Ontario L8S4M1, Canada} 
\affiliation{Canadian Institute for Advanced Research, Toronto, Ontario M5G1Z8, Canada}
\author{Bayrammurad I. Saparov}%
\author{Athena S. Sefat}%
\affiliation{Materials Science and Technology Division, Oak Ridge National Laboratory, Oak Ridge, Tennessee 37831, USA}


\date{\today}

\begin{abstract}
We take advantage of the site-selective nature of the $^{75}$As and $^{63}$Cu NMR techniques to probe the Cu substitution effects on the local magnetic properties of the FeAs planes in Ba(Fe$_{1-x}$Cu$_x$)$_2$As$_2$.  We show that the suppression of antiferromagnetic Fe spin fluctuations induced by Cu substitution is weaker than a naive expectation based on a simple rigid band picture, in which each Cu atom would donate 3 electrons to the FeAs planes.  Comparison between $^{63}$Cu and $^{75}$As NMR data indicates that spin fluctuations are suppressed at the Cu and their neighboring Fe sites in the tetragonal phase, suggesting the strongly local nature of the Cu substitution effects.  We attribute the absence of a large superconducting dome in the phase diagram of Ba(Fe$_{1-x}$Cu$_x$)$_2$As$_2$ to the emergence of a nearly magnetically ordered FeAs planes under the presence of orthorhombic distortion. 
\end{abstract}

\maketitle  
Since the discovery of iron pnictide high $T_{\rm{c}}$ superconductors \cite{Kamihara}, strong attention has been placed on its superconducting mechanism.  The superconducting phase emerges once chemical substitution or application of pressure suppresses magnetic and structural phase transitions.  This suggests that magnetic and structural instabilities, and their fluctuations, may be playing a role in Cooper paring.  In the case of the BaFe$_2$As$_2$ system, substitution of K$^+$ ions into Ba$^{2+}$ sites introduces holes into the FeAs planes, resulting in $T_{\rm{c}}$ = 38 K for 40$\%$ substitution \cite{Rotter}.  One can also substitute Co to Fe sites, and Ba(Fe$_{1-x}$Co$_x$)$_2$As$_2$ exhibits the maximum $T_{\rm{c}}$ $\simeq $ 25 K for  $x$ $ \lesssim $ 0.08 \cite{Sefat}.  Since Co is next to Fe in the atomic periodic table, the primary effects of Co substitution is generally considered to be donation of the extra electron into the FeAs planes.  The systematic shift of the Fermi energy observed by ARPES (angle-resolved photoemission spectroscopy) measurements seems to support such a rigid band picture \cite{Brouet, Neupane, Ideta}.  Furthermore, the optimum $T_{\rm{c}}$ of the Ni substituted Ba(Fe$_{1-x}$Ni$_x$)$_2$As$_2$ and Cu substituted Ba(Fe$_{1-x}$Cu$_x$)$_2$As$_2$ requires the doping level, $x$, to be smaller than the case of Co by a factor of 2$\sim$3 \cite{Canfield, Ni, Mun}.  In view of the fact that Ni and Cu possess two and three extra electrons over Fe, respectively, these results also seem to support the rigid band picture.  

It turned out, however, that disorder caused by structural defects alone may be sufficient to induce superconductivity in SrFe$_2$As$_2$ \cite{Saha}.  Moreover, density functional theory calculations suggest that the extra electrons of Co, Ni, and Cu may be strongly bound to the dopants, contrary to the expectations from a simple rigid band picture \cite{Wadati}.  More recent substitution studies also suggest that Ni and Cu substitution is not exactly 2 and 3 times more effective than Co, respectively \cite{Ni, Ideta}.  

In this study, we report $^{75}$As ($I$ = 3/2, $\gamma_n/2\pi$ = 7.292 MHz/T) and $^{63}$Cu ($I$ = 3/2, $\gamma_n/2\pi$ = 11.285 MHz/T) NMR measurements on Ba(Fe$_{1-x}$Cu$_x$)$_2$As$_2$ to reveal the Cu substitution effects on the local magnetic properties of BaFe$_2$As$_2$.  We found that Cu substitution progressively suppresses Fe spin fluctuations in the FeAs planes based on the composition $x$ dependence of the $^{75}$As nuclear spin-lattice relaxation rate divided by temperature, $^{75}(1/T_1T)$.  The observed effect of Cu substitution, however, is not as strong as expected from the simplistic rigid band model.  Furthermore, we fully take advantage of the local nature of the NMR techniques, and demonstrate that antiferromagnetic spin fluctuations are completely suppressed at the Cu site in the tetragonal phase above the tetragonal to orthorhombic structural transition at $T_s$.  Moreover, their influence spatially extends only at the FeAs sites in their vicinity.  The presence of the static orthorhombic distortion below $T_s$ transforms the non-magnetic character of the Cu site, leading to an almost magnetically ordered ground state; this finding explains why the bulk superconductivity is nearly non-existent for the Ba(Fe$_{1-x}$Cu$_x$)$_2$As$_2$ series.  In fact, none of our samples are superconducting, and only one sample with $x=0.044$ is known to exhibit a resistive superconducting transition with $T_{\rm{c}}$ as low as $\sim 2$~K \cite{Ni}. 

We grew the single crystals by the self-flux method, and determined the Cu concentration by energy dispersive x-ray analysis as $x=0.023$, 0.04, and 0.058.  Our electrical resistivity $R$ and magnetization $M$ data (see the Supplementary Material \cite{Supplement}) are consistent with earlier reports \cite{Canfield, Ni}.  Fig.~1 is the summary of the bulk properties of Ba(Fe$_{1-x}$Cu$_x$)$_2$As$_2$ in the phase diagram adopted from \cite{Ni}.  We used our $R$ and $M$ data \cite{Supplement} to determine $T_{s} \sim 90$~K and spin density wave (SDW) transition temperature, $T_{SDW} \sim 83$~K, for $x=0.023$, while our X-ray diffraction data showed $T_{s} = 47 \pm 2$~K for $x=0.04$ (see below).  Interpolation of the results from \cite{Ni} suggests $T_{s} = 10 \sim 20$~K for $x=0.058$, which is consistent with the resistivity upturn observed in our crystal \cite{Supplement}.

\begin{figure}
\includegraphics[width=3in]{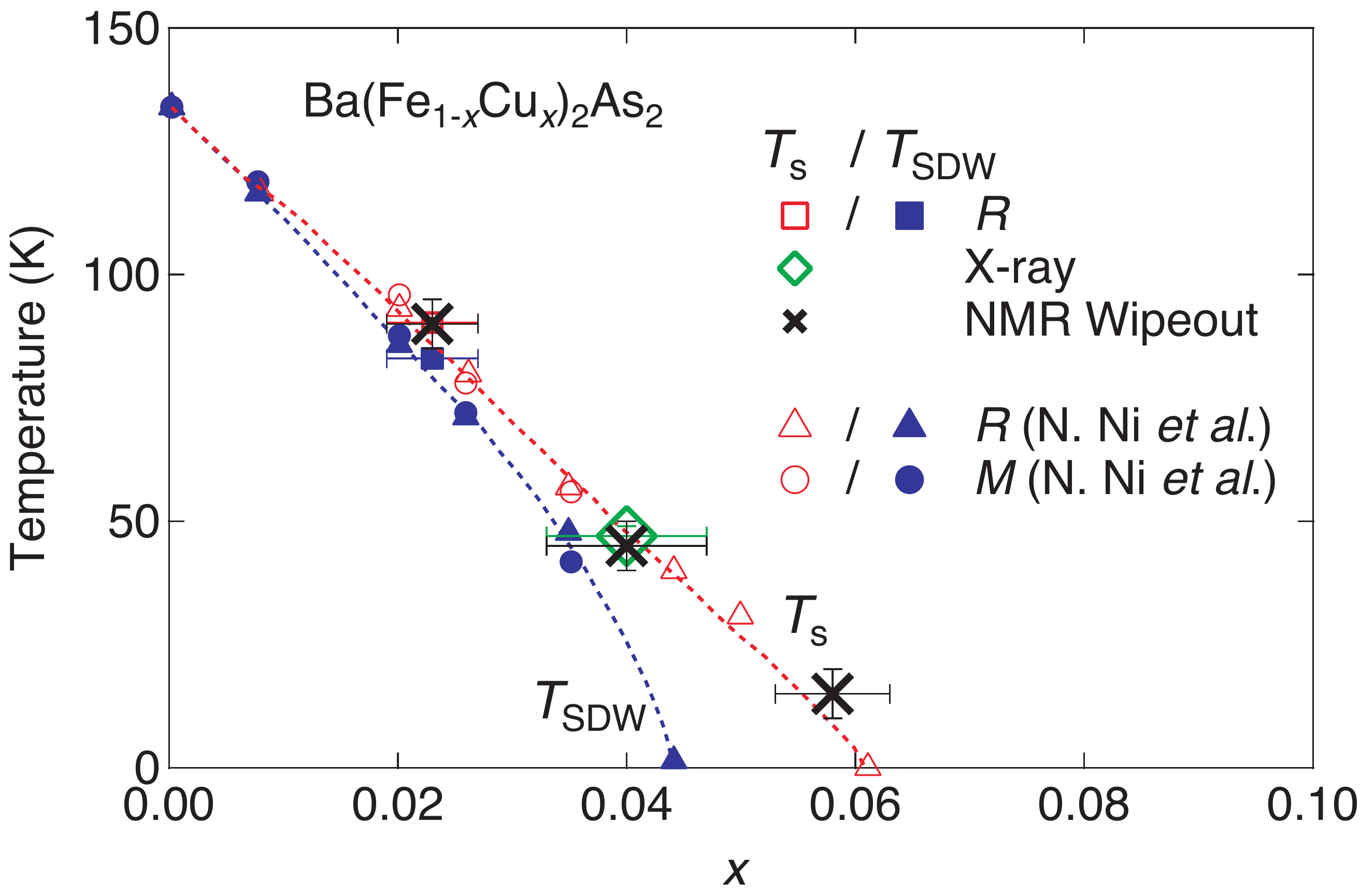}
\caption{\label{fig:R_MH} (Color online) The phase diagram of Ba(Fe$_{1-x}$Cu$_x$)$_2$As$_2$ for the structural ($T_s$) and Spin Density Wave ($T_{SDW}$) transitions reported by N.~Ni et al. \cite{Ni}, together with the results for our crystals with $x=0.023$ and $x=0.04$.  Also shown is the onset temperature of the NMR signal intensity wipeout ($\times$) as determined from Fig.~3(c); notice the clear correlation with $T_s$.}
\end{figure}

We show representative $^{75}$As NMR spectra in Fig.~2(a).  We observed a set of strong signals composed of a center line ($I_z$ = $1/2$$\leftrightarrow$$-1/2$ transition) at $\sim57.1$~MHz and two quadrupole-split satellite lines ($I_z$ = $\pm1/2$$\leftrightarrow$$\pm3/2$ transitions) at $\sim 54.4$~MHz and $\sim 59.8$~MHz.  These peaks correspond to the NMR signals observed for the undoped FeAs planes of BaFe$_2$As$_2$ \cite{Kitagawa}, and arise from the {\it host} As$_{\rm{host}}$ sites with no Cu in their neighbors.  The quadrupole splitting $\nu_c$ between the center and satellite lines for the As$_{\rm{host}}$  sites, which is governed by the local structural environment through the nuclear quadrupole interactions, shows a slight decrease with increasing $x$.  

We also observed a hump at $\sim$57.12 MHz near the center line of As$_{\rm{host}}$ (see the main panel of Fig.~2(a)) and the corresponding quadrupole split satellite peaks at $\sim$53.8 MHz and $\sim$60.4 MHz (see the inset of Fig.~2(a)).  Judging from the relatively large spectral weight, we assign these signals to the  As$_{\rm{n.n.n.}}$ sites, whose next nearest neighbor Fe sites is substituted by Cu (see Fig.~2(b) for the geometrical configuration).  We found that $^{75}(1/T_1)$ at As$_{\rm{n.n.n.}}$ site is comparable to, but somewhat slower than, that observed at As$_{\rm{host}}$ site, as shown in Fig.~6(c) below, suggesting that the local magnetic properties at the As$_{\rm{n.n.n.}}$ and As$_{\rm{host}}$ sites may be qualitatively similar.  

Besides the NMR signals arising from the As$_{\rm{host}}$ and As$_{\rm{n.n.n.}}$ sites, we observed a small peak at $\sim$55 MHz as shown in the inset of Fig.~2(a).  The intensity of this peak grows with $x$.  Since we found no corresponding signals on the higher frequency side of the center peak of the As$_{\rm{host}}$ sites, we cannot attribute this peak to a satellite transition split by the first order quadrupole effect.  Instead, we assign this peak to the center line of the nearest neighbor of Cu, the As$_{\rm{n.n.}}$ site, in analogy with the case of the Co substituted samples \cite{Ning4}.  Since the Cu substitution generates stronger lattice strain \cite{Canfield, Ni}, the second order quadrupole splitting between the center lines of As$_{\rm{host}}$ and As$_{\rm{n.n.}}$, $\sim$2.1 MHz, is larger than that observed for the Co substituted samples, $\sim$0.4 MHz \cite{Ning1, Ning4}.   

\begin{figure}[t]
\includegraphics[width=8.5cm]{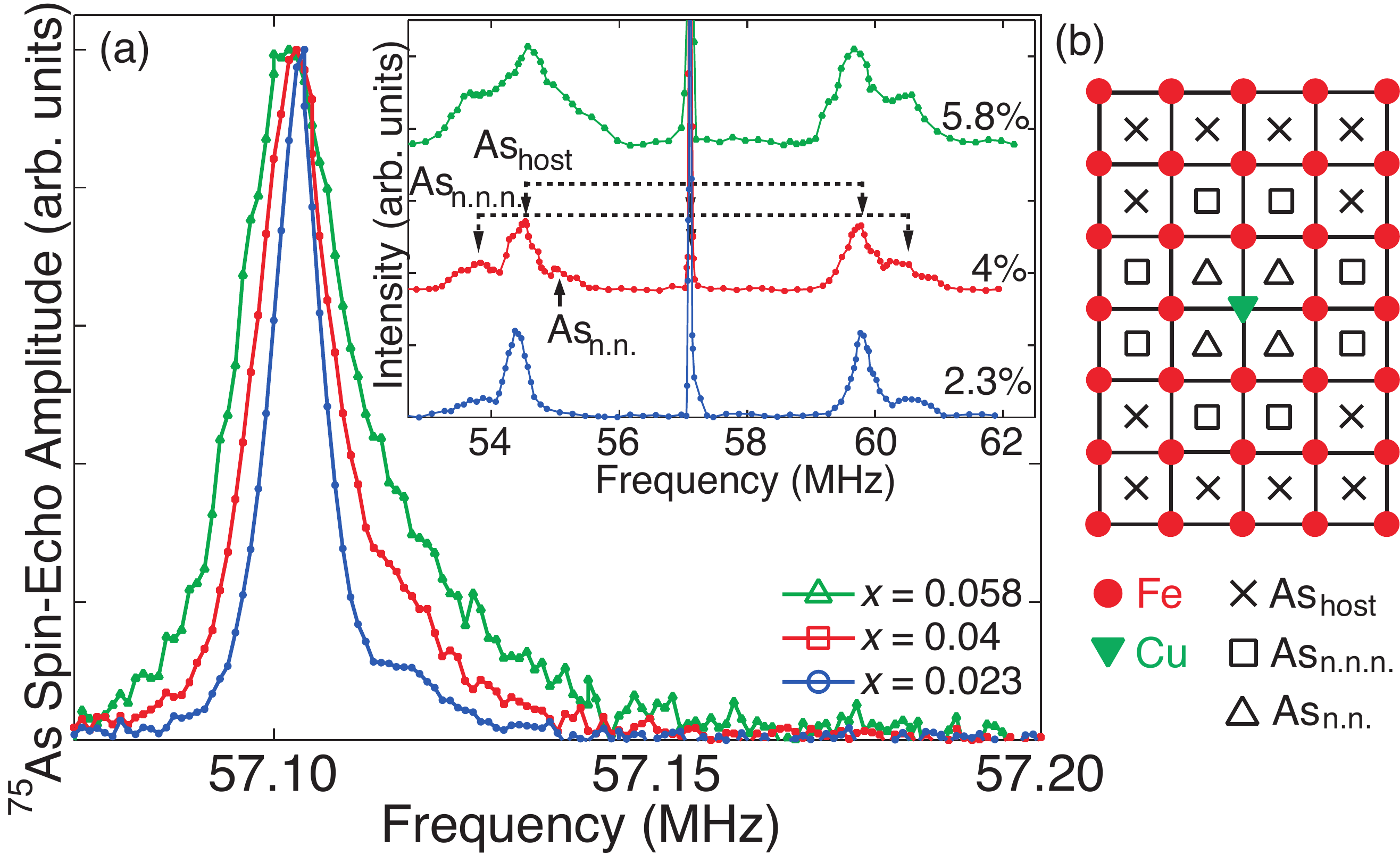}
\caption{\label{fig:250K_spc} (Color online) (a)  Fourier-transformed $^{75}$As NMR spectra of the center lines for Ba(Fe$_{1-x}$Cu$_x$)$_2$As$_2$ at 250 K with the external magnetic field $H$ = 7.8076 T applied along the $c$-axis. The inset shows the satellite lines.  (b) A schematic picture of the environment around the As$_{\rm{host}}$ ($\times $), As$_{\rm{n.n.n.}}$ ($\Box$) and As$_{\rm{n.n.}}$ ($\bigtriangleup $) sites.}
\end{figure}

\begin{figure}[t]
\includegraphics[width=3in]{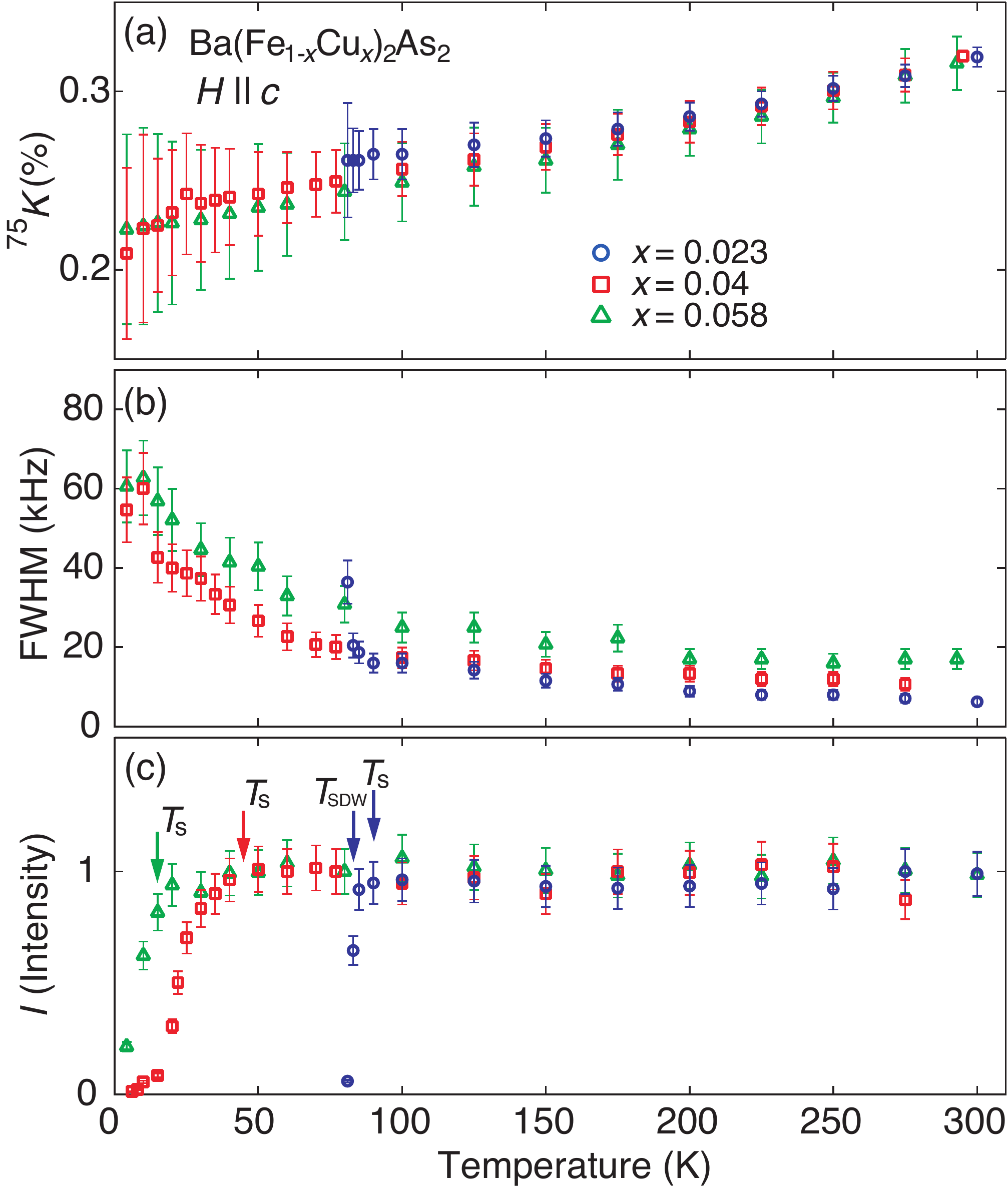}
\caption{\label{fig:K_nu_q} (Color online) (a) $^{75}K$, (b) FWHM for the center line, and (c) the integrated NMR intensity $I$ (normalized by the Boltzmann factor) for Ba(Fe$_{1-x}$Cu$_x$)$_2$As$_2$.}
\end{figure}

In Fig.~3(a), we summarize the $T$ and $x$ dependences of the local magnetic susceptibility of the host FeAs plane as determined from the $^{75}$As NMR Knight shift of the As$_{\rm{host}}$ site.  Upon cooling, $^{75}K$ decreases monotonically.  The higher the Cu concentration, $^{75}K$ is slightly smaller.  These trends are analogous to the case of Co substitution \cite{Ning4}.  The full width at half maximum (FWHM) of the center line of the spectra increases with decreasing $T$ as summarized in Fig.~3(b).  In $x$ = 0.023, we observed an abrupt onset of the increase of the FWHM below $T_{\rm{s}}\sim 90$~K, accompanied by a strong upturn of $^{75}(1/T_1T)$ as shown in Fig.~4.  These results suggest that the static orthorhombic distortion below $T_{\rm{s}}\sim 90$~K prompts the growth of the short range SDW order.  These anomalies are followed by the complete disappearance of the paramagnetic NMR signals below $T_{\rm{SDW}}\sim 83$~K.  NMR measurements in the SDW ordered state below $T_{\rm{SDW}}$ is beyond the scope of the present work. 

The strong anomalies at $T_{\rm{s}}$ and $T_{\rm{SDW}}$ in the temperature dependence of $R$, $M$, and FWHM observed for the $x=0.023$ sample are absent for our $x$ = 0.04 and 0.058 samples.  We found, however, that the integrated NMR intensity $I$ (normalized by the Boltzman factor) begins to be wiped out below $\sim$45 K and $\sim$15 K for $x$ = 0.04 and 0.058, respectively, as shown in Fig.~3(c).  The onset temperature of the NMR signal wipeout plotted in the phase diagram in Fig.~1 ($\times$ symbols) suggests a correlation with the structural phase transition at $T_s$.  

\begin{figure}[t]
\includegraphics[width=3in]{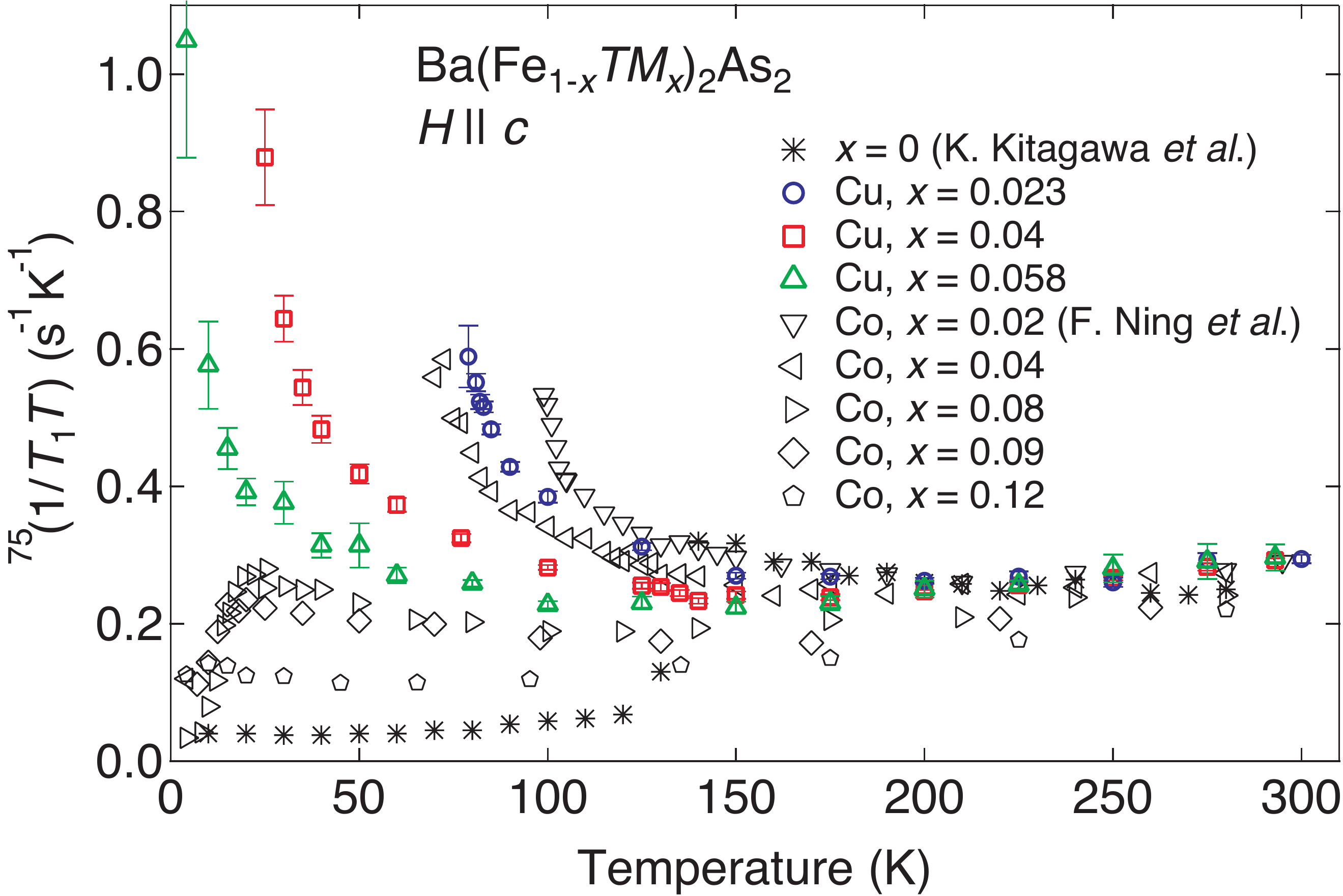}
\caption{\label{fig:T1T} (Color online) $^{75}(1/T_1T)$ observed for Ba(Fe$_{1-x}$Cu$_x$)$_2$As$_2$ and Ba(Fe$_{1-x}$Co$_x$)$_2$As$_2$ (from \cite{Ning2, Kitagawa}).}
\end{figure}

\begin{figure}[t]
\includegraphics[width=3in]{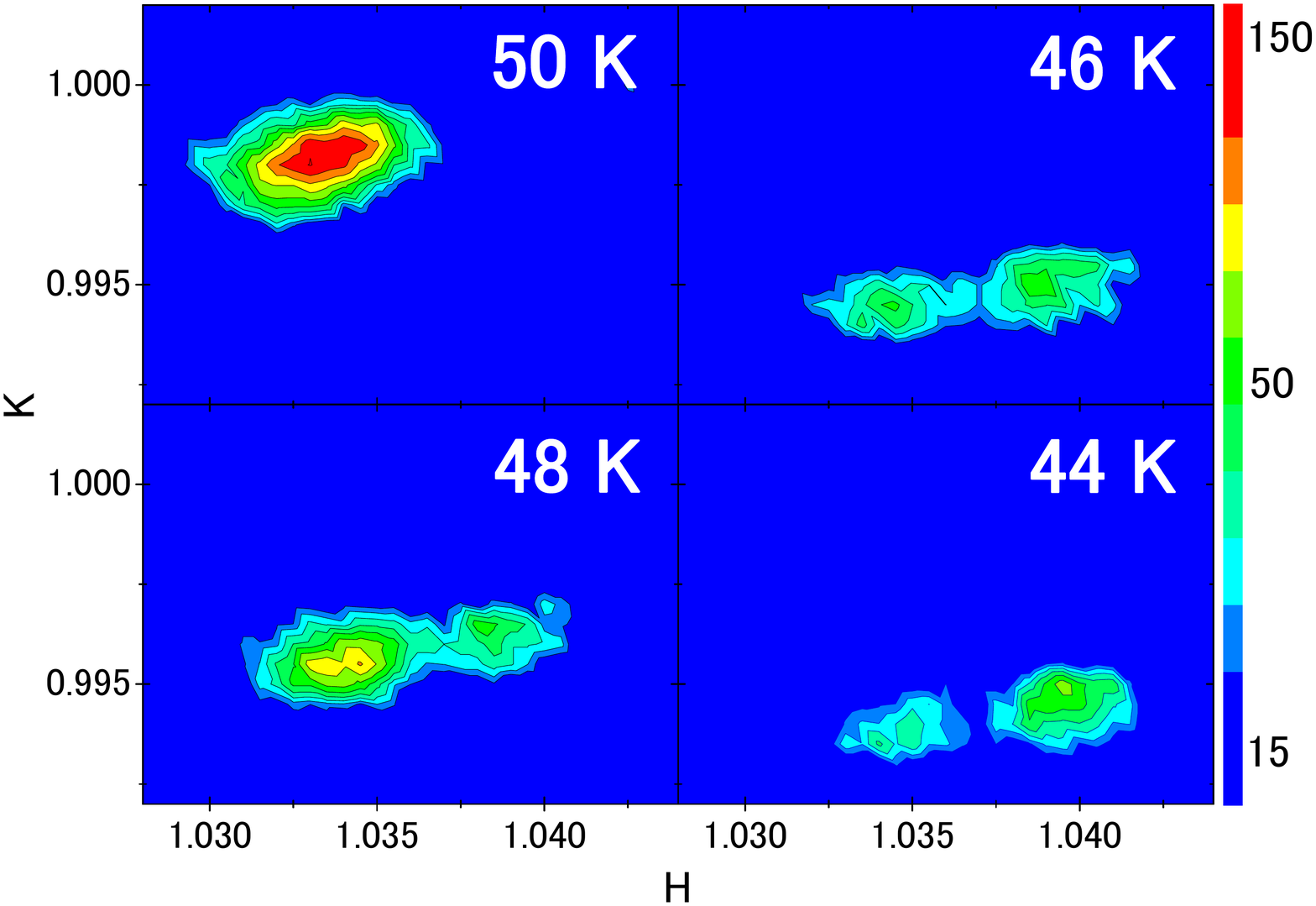}
\caption{\label{fig:X-ray} (Color online) X-ray diffraction reciprocal space maps within the HK plane near (1,1,6) observed for Ba(Fe$_{1-x}$Cu$_x$)$_2$As$_2$ with $x=0.04$.  The splitting of the intensity in the H-direction below 50 K indicates the tetragonal to orthorhombic phase transition near $T_{\rm{s}}=47\pm 2$~K.}
\end{figure}

In order to definitively test the effects of the structural phase transition, we performed X-ray scattering measurements for the same piece of $x$ = 0.04 crystal used for our NMR measurements.  We used Cu- K$_{\alpha 1}$ radiation ($\lambda =1.54041$ \AA  ) produced by an 18 kW rotating anode x-ray source with a perfect germanium (111) monochromator. We mounted the crystal on the coldfinger of a closed-cycle helium cryostat and aligned within a four-circle Huber diffractometer. The temperature of the sample was maintained to within $\pm$0.1~K. X-ray measurements primarily focused on the (1,1,6) HTT Bragg peak, as indexed using the tetragonal notation of the high-temperature tetragonal (HTT) phase.  Measurements of the H-K plane at L=6 and for temperatures between 44 K and 50 K are shown in Fig.~\ref{fig:X-ray}.  This particular set of measurements was performed on cooling, and shows a single HHT peak split into two orthorhombic twins, primarily along the H direction.  Other independent warming and cooling runs show similar results, but with some history dependence to both the intensities of the orthorhombic twin scattering, and the temperature at which the transition occurs, consistent with a discontinuous phase transition at $T_{\rm{s}}=47 \pm 2$~K.  

The comparison of the NMR data in Fig.~3(c) and X-ray scattering results suggests that this phase transition at $T_{\rm{s}}$ is not a trivial structural phase transition, unlike the case of the high temperature tetragonal to orthorhombic structural phase transition in high $T_{\rm{c}}$ cuprates La$_{2-x}$Sr$_{x}$CuO$_{4}$.  In the latter, the structural transition has no noticeable effect on low frequency spin dynamics within the CuO$_2$ planes, and the NMR signal intensity exhibits no anomaly in the orthorhombic phase \cite{imai1993}.  In contrast, the loss of the $^{75}$As NMR signal intensity below $T_s$ in Fig.~3(c) in the present case indicates that the NMR relaxation rates become divergently fast in the non-observable domains of the crystal, which implies that the orthorhombic distortion prompts Fe spin fluctuations to slow down rather abruptly by enhancing the short range SDW order \cite{Fu}.  The latter is also evidenced by the upturn of $^{75}(1/T_{1}T)$ in Fig.~4 for the still observable segments of the sample.  Taken together, our NMR results indicate that Fe spins are on the verge of a magnetic phase transition in orthorhombic domains below $T_s$, even though the magnetic susceptibility data may not give a definitive hint of a magnetic phase transition.  We emphasize that Fe spins are {\it not} uniformly ordering below $T_s$, because the existence of both the observable and non-observable NMR signals implies a huge distribution of $^{75}(1/T_{1}T)$, and hence inhomogeneous nature of the slowing down of Fe spin fluctuations.  That is, Fe spins are progressively freezing in a glassy manner.  This conclusion is also consistent with our finding that $^{75}(1/T_{1}T)$ begins to develop a mild distribution below $\sim 100$~K even for the observable NMR signals.
   
In Fig.~4, we compare the concentration dependence of $^{75}(1/T_1T)$ measured at the center line of the As$_{\rm{host}}$ site with our earlier report for the Co-substituted samples \cite{Ning1, Ning2, Ning4}.  In $x$ = 0.023, $^{75}(1/T_1T)$ shows a monotonic increase below $\sim$150 K due to a gradual growth of short range SDW order toward the SDW transition at $T_{\rm{SDW}}$ $\sim 83$~K.  With increasing the Cu concentration, the enhancement of $^{75}(1/T_1T)$ is progressively suppressed.  At a qualitative level, our new results are similar to the concentration dependence previously reported for the Co \cite{Ning1, Ning2, Ning4} or Ni \cite{Zhou, Dioguardi} substituted BaFe$_2$As$_2$.  

Compared with Ba(Fe$_{1-x}$Co$_x$)$_2$As$_2$, we found that Cu substitution suppresses $^{75}(1/T_1T)$ more strongly.  This is consistent with the fact that the SDW phase transition is suppressed by a smaller amount of Cu substitution than Co \cite{Canfield, Mun, Ni}.  In a simplistic rigid band model, we expect that the substituted Cu introduces 3$x$ electrons, while Co dopes $x$ electrons to the FeAs planes.  Our results are clearly inconsistent with such expectations, because $^{75}(1/T_1T)$ for $x$ = 0.023 is not as strongly suppressed as that for Ba(Fe$_{0.93}$Co$_{0.07}$)$_2$As$_2$.  This finding indicates that the rigid band model alone is not sufficient to account for the composition dependence of the electronic properties of Ba(Fe$_{1-x}$TM$_x$)$_2$As$_2$ (TM=Co and Cu).  

\begin{figure}[t]
\includegraphics[width=8.5cm]{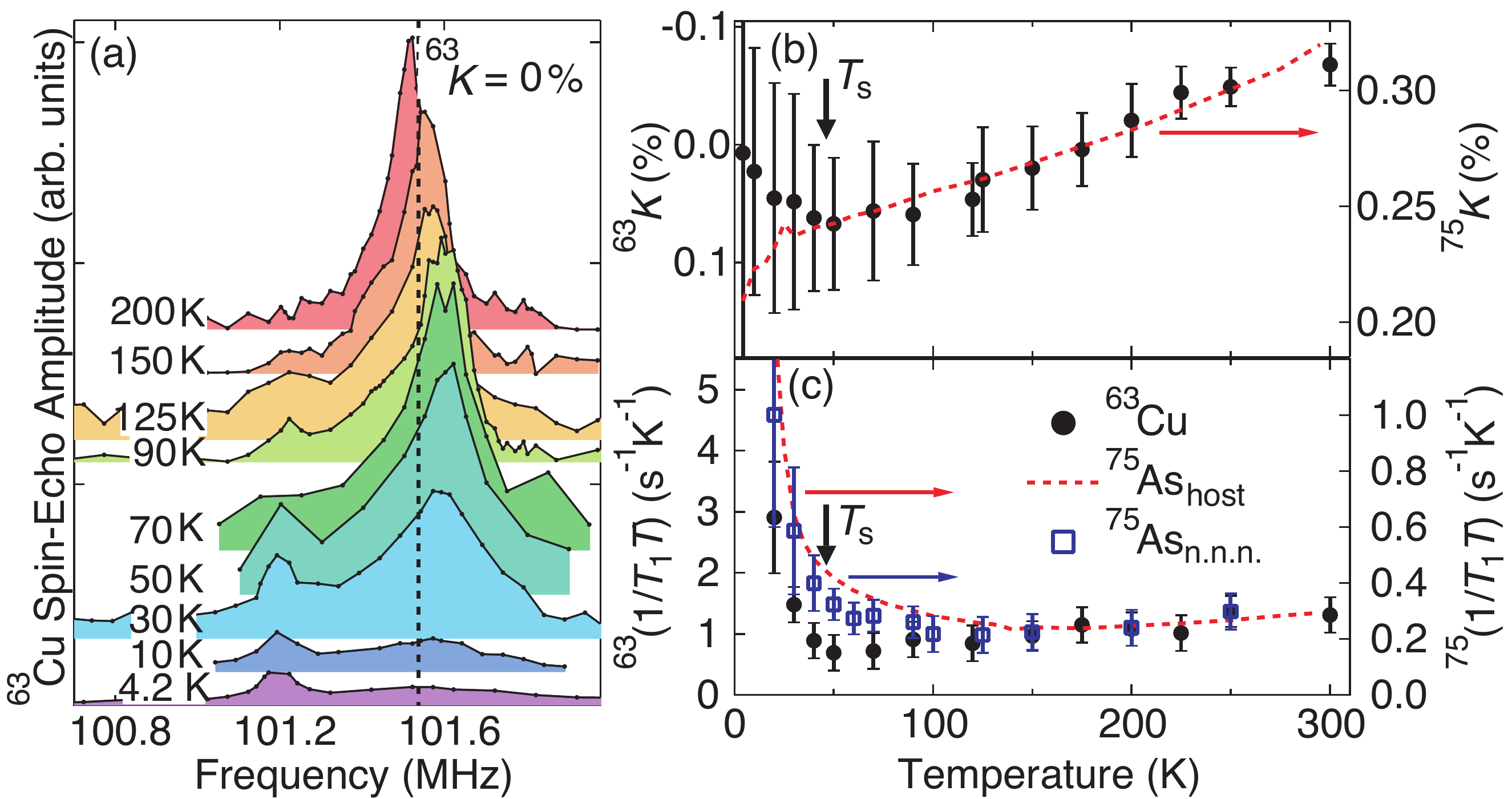}
\caption{\label{fig:Cu_NMR} (Color online) (a) Temperature dependence of $^{63}$Cu NMR spectra observed for Ba(Fe$_{0.96}$Cu$_{0.04}$)$_2$As$_2$ in $H$ = 8.9976 T applied along the $c$ axis.  The signal intensity is normalized for the Boltzmann factor.  (b) $^{63}K$ and $^{75}K$.  (c) $^{63}(1/T_1T)$ and $^{75}(1/T_1T)$.  $^{75}(1/T_1T)$ at the As$_{\rm{host}}$ site was measured using the center peak.  $^{75}(1/T_1T)$ at the As$_{\rm{n.n.n.}}$ was measured using the satellite peak to avoid contamination by the strong signal from the As$_{\rm{host}}$ site.}
\end{figure}

NMR measurements conducted at the $^{63}$Cu sites of the $x$ = 0.04 sample provide additional and clear insight into the nature of the Cu substitution effects.  We present $^{63}$Cu NMR spectra in Fig.~6(a).  Upon cooling, the spectrum shifts to higher frequency down to $T_{\rm{s}}$, scaling with the $T$ dependence of $^{75} K$, as shown in Fig.~6(b).  Below $T_{\rm{s}}$, the spectrum is gradually broadened and wiped out.  The small peak observed at $\sim101.2$~MHz does not exhibit any shift with decreasing $T$ and is probably from a very small amount of an impurity phase.  

Fig.~6(c) summarizes $^{63}(1/T_1T)$ measured at the $^{63}$Cu sites.  While $^{75}(1/T_1T)$ measured at the As$_{\rm{host}}$ site begins to increase below $\sim$150 K due to the growth of short range SDW, $^{63}(1/T_1T)$ monotonically decreases down to $T_{\rm{s}}$.  We conclude that antiferromagnetic spin fluctuations are {\it locally} suppressed around the Cu sites in the tetragonal phase.  It is also important to realize that the temperature dependence of $^{75}(1/T_1T)$ measured at the As$_{\rm{n.n.n.}}$ site is in between that of Cu and As$_{\rm{host}}$ sites below $\sim 100$~K.  This means that the influence of the substituted Cu atoms extends to the n.n. Fe sites, and possibly to the n.n.n. Fe sites as well.  Notice that if the n.n. and n.n.n. Fe sites behave the same way as the host Fe sites that bond with As$_{\rm{host}}$ sites,  $^{75}(1/T_1T)$ at the As$_{\rm{n.n.n.}}$ would show identical temperature dependence as that at the As$_{\rm{host}}$ site.  Analogous contrasting behavior between Fe and Co spin dynamics was previously observed for Ba(Fe$_{1-x}$Co$_{x}$)$_2$As$_2$ \cite{Ning1, Ning3}, but the spatial extension of the influence of Co was not investigated.  
 
Our NMR data provide clear evidence that, in the tetragonal phase above $T_{\rm{s}}$, substituted Cu atoms significantly alter the electronic properties of FeAs planes in their vicinity, rather than merely donating $3x$ electrons uniformly to FeAs planes.  On the other hand, $^{63}(1/T_1T)$ begins to increase below $T_{\rm{s}}$.  Furthermore, the $^{63}$Cu NMR signal intensity decreases below $T_{\rm{s}}$ as shown in Fig.~6(a), concomitant with the wipeout of the $^{75}$As NMR signals shown in Fig.~3(c).  These findings indicate that the structural transition into the orthorhombic phase promotes strong spin fluctuations even at the Cu sites and their vicinity.  It is not clear why apparently non-magnetic Cu sites in the tetragonal phase change their character below $T_{\rm{s}}$.  Perhaps the nature of hybridization of the Cu 3d orbitals with neighboring Fe sites changes under the presence of orthorhombic distortion, and some Cu 3d orbitals which gain extra weight in the orthorhombic phase help the growth of the SDW correlations.  Alternately, orthorhombic domains may pin the SDW domain boundaries \cite{Mazin}, resulting in strong enhancement of low frequency spin fluctuations within the static domains.       

To summarize, we conducted $^{75}$As and $^{63}$Cu NMR measurements on Ba(Fe$_{1-x}$Cu$_x$)$_2$As$_2$, and investigated the Cu substitution effects on the local magnetism of FeAs planes with a primary focus on $x=0.04$.  By taking advantage of the local nature of the NMR techniques, we demonstrated that Cu sites behave as non-magnetic defects in the tetragonal phase, but the orthorhombic distortion below $T_{\rm{s}}$ leads to magnetic behavior of the Cu sites.  Our NMR data establish that the influence of the substituted Cu atoms on the magnetic properties of FeAs planes is more localized than generally believed, and a simplistic rigid band picture is insufficient to account for the properties of Ba(Fe$_{1-x}$TM$_x$)$_2$As$_2$.  Our observation of glassy spin freezing of FeAs planes under the presence of orthorhombic distortion below $T_s$ also explains why the superconducting dome in the phase diagram of Ba(Fe$_{1-x}$Cu$_x$)$_2$As$_2$ \cite{Ni} is almost entirely suppressed.

The work at McMaster was supported by NSERC and CIFAR.  H.T. was supported by the Grant-in-Aid for Japan Society for the Promotion of Science Fellows during his visit at McMaster, and M.T. is on a leave from NIMS, Japan.  Research at Oak Ridge National Laboratory was supported by U.S. Department of Energy, Basic Energy Sciences, Materials Sciences and Engineering Division.  \\

\begin{figure}
\includegraphics[width=6cm]{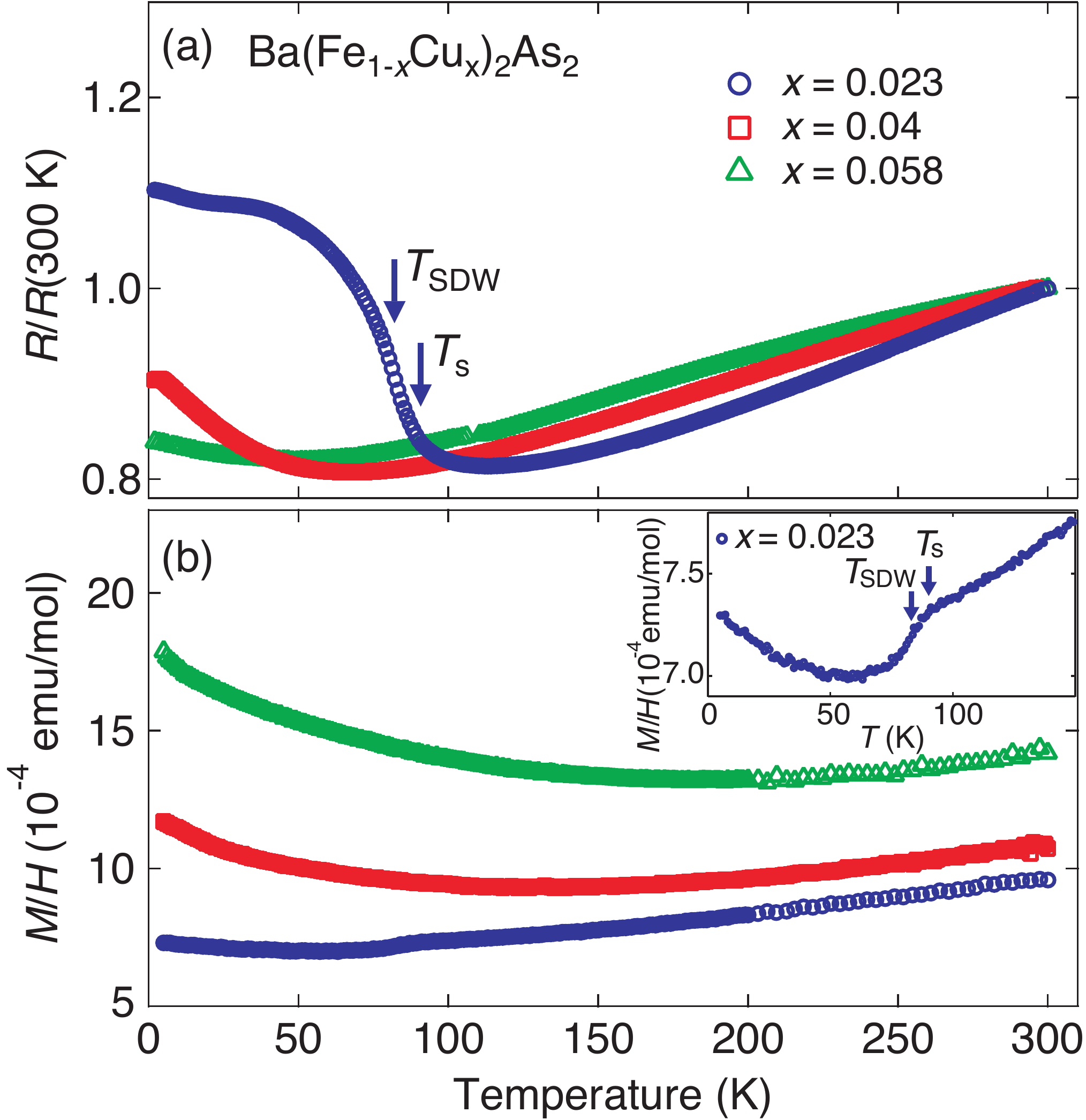}
\caption{\label{fig:Cu_NMR} (a) The electrical resistivity $R$ measured for the Ba(Fe$_{1-x}$Cu$_{x}$)$_{2}$As$_{2}$ single crystals.  We normalized $R$ with the room temperature value, $R$(300K).  The discontinuous tetragonal to orthorhombic structural phase transition is located at $T_{s} \sim$ 47~K for the x = 0.04 sample, according to our X-ray diffraction measurements; notice the upturn of $R$ below $T_{s}$.  Analogous upturn of $R$ is observed also for the $x = 0.058$ sample near the expected $T_{s} \sim$ 15 K.  (b) The magnetization $M$ of the Ba(Fe$_{1-x}$Cu$_{x}$)$_{2}$As$_{2}$ single crystals measured in $H = 1$~T.}
\end{figure}

\begin{figure}
\includegraphics[width=6cm]{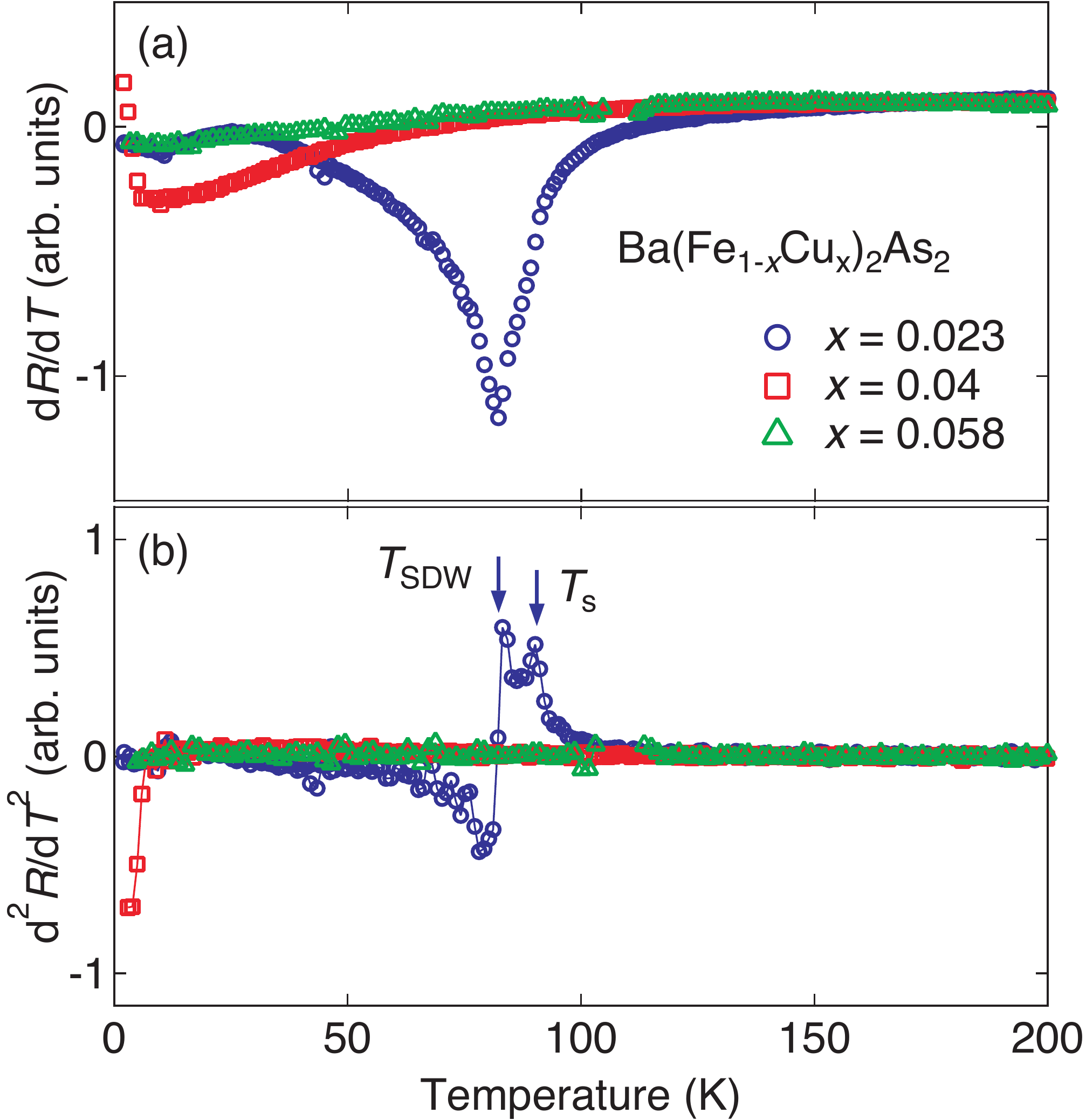}
\caption{\label{fig:Cu_NMR} (a) First and (b) second derivatives of $R$.  The signatures of the successive structural and SDW transitions at $T_{s} \sim 90$~K and $T_{SDW}\sim 83$~K are clearly seen for the $x = 0.023$ sample. }

\end{figure}


\end{document}